\documentclass{phbauth}
\usepackage{graphicx}

\begin{document}

\begin{frontmatter}

\title{Transport measurements in BSCCO-2212 in high magnetic fields}

\author[address1]{J.M. Keartland\thanksref{thank1}},
\author[address2]{R.A. Doyle}

\address[address1]{Department of Physics and Condensed Matter Physics Research Unit,
University of the Witwatersrand, P.O. Wits 2050, Johannesburg,
South Africa}
\address[address2]{IRC in Superconductivity, University of Cambridge, Cambridge, United Kingdom}

\thanks[thank1]{Corresponding author. Fax: +27 11 339 8262, E-mail: ktland@physnet.phys.wits.ac.za}

\begin{abstract}
We present results for the in-plane resistivity as a function of
temperature for clean single crystal samples of BSCCO-2212 in
applied magnetic fields above 0.5 T.  Several fixed angles between
the magnetic field and the crystal $c$-axis were studied. The data
were analysed in terms of the vortex-glass (VG) model, as
suggested by previous experimental studies. Several features of
the data were found to be inconsistent with the VG scenario.
Recent theoretical studies have challenged the VG interpretation,
and these are discussed in terms of our data.
\end{abstract}


\end{frontmatter}

% The main text begins here. The \section commands are optional.

The vortex-matter-phase-diagram (VMPD) in the high temperature
cuprate superconductors (HTCS) has been the source of much
interest and controversy since their discovery.  More recently
this interest has focussed on the thermodynamic first order
transition (FOT) in single crystal samples.  In un-twinned YBCO
the situation seems to have been clarified - the FOT persists up
to the multicritical point, and above this point Arrhenius
behaviour (AB) is observed. In contrast, the highly anisotropic
HTCS, BSCCO-2212, clean single crystal samples show evidence for
the FOT at low applied magnetic fields, but it is far from clear
what occurs in the region of the VMPD above the second peak field
($B_\mathrm{sp}$).  Safar {\em et al\/} \cite{SEA}, on the basis
of electrical transport measurements, have suggested that a
thermodynamic second order transition to a vortex-glass (VG) state
occurs above 2.0 T when the applied field is parallel to the
crystal $c$-axis.  This conclusion was based on the observation
that departures from AB occur in the tail of the $R$ versus $T$
curve in this field regime, and that the resistance followed the
VG scaling law
\begin{equation}
R \propto (T - T_\mathrm{g})^n. \label{e.TDGL}
\end{equation}
where $T_\mathrm{g}\ $ is the proposed VG transition temperature,
and $n = \nu (z - 1)$ is the critical exponent. In order to
further our understanding of what occurs at high magnetic fields
we have made measurements of the in-plane resistivity in the high
magnetic field region as a function of the temperature for several
angles of the applied field with respect to the crystal $c$-axis.
The data have been analysed in terms of (\ref{e.TDGL}).

Three narrow strip-shaped crystals of BSCCO (grown by the
travelling solvent floating zone method) were selected from a
batch of slightly over–doped single crystal samples. Gold wires
were attached to the sides and surface of the crystals for the
usual four–point measurement of the $ab$ –-plane resistance.
Nanovolt resolution was obtained using lock-in amplifiers and low
noise transformers.  Magnetic fields above 0.5 T were obtained
using a superconducting magnet operating in the persistent mode.
$R_\mathrm{ab}$ vs $T$ measurements at low applied magnetic fields
showed the characteristic "melting" step seen in other clean
crystals, confirming the high quality of the crystals.  All data
were taken at fields of 0.5 T and above, where the effects of
surface barriers are small.

The reactive background was carefully subtracted from all curves
before analysis of the data.  Arrhenius plots of $\ln
R_\mathrm{ab}\ $ vs $T^{-1}$ for all data sets above 0.5 T showed
clear departures from AB in the tail of the curve. This departure
was more marked when $B_0$ was applied at angles of 60$^\circ$ and
75$^\circ$ to the crystal $c$-axis.  The data were analyzed using
the method of \cite{SEA}. We plot $( \mathrm{d}\ \ln R/
\mathrm{d}\ T)^{-1}$ vs $T$, so that any region in the transition
which obeys (\ref{e.TDGL}) appears as a linear region in the
construction. A linear fit of this region enables us to obtain
$n$, $T_\mathrm{g}\ $ and the onset temperature $T^*$ as defined
by \cite{SEA}.  A plot of $n$ for all angles and fields is shown
in Fig. 1. In the inset to this figure we have plotted results for
$T_\mathrm{g}\ $. The results for $B_0
\parallel c$ are identical, within experimental error, to those of
\cite{SEA} in the range 2.0 T - 7.0 T. Moreover, the departure
from AB determined from our analysis is within the experimental
error of $T^*$.  The above results and discussion would appear to
support a VG scenario in BSCCO.  If this were the case we might
expect the angular behaviour to obey scaling of the form $B_0 \cos
\theta$ (the component of the magnetic field parallel to $c$), as
occurs in the case of the FOT \cite{aft}. We turn our attention to
the measurements taken at $\theta = 60^\circ$ and $\theta =
75^\circ$ as summarized in Fig. 1. The results for $n$ indicate
that there are departures from this scaling law. In an inset to
Fig. 1 we show the phase diagram for $T_\mathrm{g}\ $, from which
it may be seen that $T_\mathrm{g}\ $ does not scale with $B_0 \cos
\theta$.

\begin{figure}[t]
%h=here, t=top, b=bottom, p=separate figure page
\begin{center}\leavevmode
\includegraphics[width=0.9\linewidth]{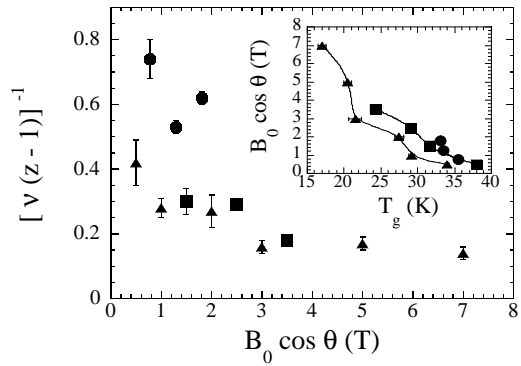}
\caption
{A summary of the results for $\nu (z-1)$ and
$T_\mathrm{g}\ $ for all angles and fields.  The triangles are for
$\theta = 0^\circ$, the squares for $\theta = 60^\circ$ and the
circles for $\theta = 75^\circ$.  The lines in the inset are to
aid the eye.}
\label{figurename}\end{center}\end{figure}

We have repeated the measurements of Safar {\em et al\/}
\cite{SEA} on high quality crystals of BSCCO (where a clear
melting step is observed) to check for VG behaviour at applied
magnetic fields above $B_\mathrm{sp}$.  While the results for $B_0
\parallel c$ are almost identical to those obtained
by \cite{SEA}, the angular dependence of the VG exponents and
temperature do not scale with $B_0 \cos \theta$, making it less
likely that the VG scenario completely accounts for the observed
data. Further work is required to understand the detailed nature
of the vortex matter in the low temperature-high field regime.

\begin{ack}
We gratefully acknowledge T. Tamegai and S. Ooi who provided the
single crystal samples used in this work.  One of us (JMK)
acknowledges financial support from the Foundation for Research
Development, South Africa, that allowed this work to be done at
the IRC in Superconductivity, and thanks the staff and students of
the IRC for their support and hospitality during his visit.
\end{ack}

\end{document}